\documentclass[aps,prb,twocolumn,superscriptaddress,showpacs,english]{revtex4-1}

\usepackage[T1]{fontenc}
\usepackage[latin9]{inputenc}
\usepackage{babel}
\usepackage{amsmath}
\usepackage{amssymb}
\usepackage{wasysym}
\usepackage{graphicx}
\usepackage{xcolor}

\definecolor{mydarkgreen}{rgb}{0.0,0.5,0.0}
\definecolor{friebrick}{rgb}{0.698,0.1333,0.1333}

\usepackage[linktocpage=true,
  colorlinks=true, 
  pdfborder={0 0 0},
  linkcolor=blue,
  citecolor=friebrick,
  filecolor=yellow,
  urlcolor=mydarkgreen,
  bookmarks,
  pdfauthor={},
]{hyperref}

\newcommand{\mpi}{Max-Planck Institut f\"ur Microstrukture Physics, Weinberg 2, 06120 Halle, Germany}
\newcommand{\cac}{CaC$_6$}
\newcommand{\mus}{$\mu^*$}
\newcommand{\mx}{MX$_2$}
\newcommand{\src}{SrC$_2$}
\newcommand{\rg}{RbGe$_2$}
\newcommand{\rs}{RbSi$_2$}
\newcommand{\tc}{T$_{\textmd C}$}
\newcommand{\omlog}{$\omega_{\textmd log}$}
\newcommand{\dos}{$N(E_{{\textmd F}})$}  

\begin{document}

\title{Superconductivity in intercalated group-IV honeycomb structures}

\author{Jos\'e A. Flores-Livas}
\author{Antonio Sanna}
\affiliation{\mpi}
\date{\today}

\begin{abstract}
We present a theoretical investigation on electron-phonon superconductivity of honeycomb \mx\ layered structures.
Where X is one element of the group-IV (C, Si or Ge) and M an alkali or an alkaline-earth metal.
Among the studied composition we predict a \tc\ of 7\,K in \rg, 9\,K in \rs\ and 11\,K in \src. 
All these compounds feature a strongly anisotropic superconducting gap.
Our results show that despite the different doping level and structural properties, 
the three families of materials fall into a similar description of its superconducting behavior.
This allows us to estimate an upper critical temperature of about 20\,K for the class of intercalated group-IV structures, 
including intercalated graphite and doped graphene. 
\end{abstract}

\pacs{~}

\maketitle

A large research effort has been lately focused on atomic-thin layered materials and their 
properties~\cite{Geim_NatureMaterials2007,Geim_Science2009,Novoselov_Nature2012}. 
This was triggered by the creation of graphene from graphite~\cite{Novoselov_graphene_Science2004} and also motivated  
by the belief in many potential applications since thin systems can be significantly modified in their electronic properties 
simply by acting on parameters as stacking, chemical and physical doping~\cite{Fedorov_Nat2014,Yang_Nat2014}. 
In fact this versatility is an extraordinary playground for searching for new superconductors (SC)~\cite{Layered_Superconductors}. 
Many (low temperature) SC are already known in the class of graphite intercalated compounds 
GICs~\cite{Belash_LiC2_SSC1989,Nalimova_LiGICs_Carbon1995,Avdeev_AlkaliGICs_HiPRes1990,Weller_CaC6_NAT2005,Nemery_PRL_CaC6}, graphene 
itself has been predicted to superconduct with a critical temperature (\tc) of 18\,K upon Li doping~\cite{Profeta_Graphene_Nat2012}.

Among all possible compounds, those chemically and structurally closer to graphite are the honeycomb lattices of 
silicon~\cite{Sanfilippo_PRB,Bordet_PRB_CaSi2,Imai_BaSi2_hiPT_PRB1998,MImai_ChemMat} and germanium~\cite{Tobash_CaGe2_JSSChem2007,Zintl_EuGe2_JSSChem2004}. 
For which suplerconductivity upon intercalation was also reported~\cite{Sanfilippo_PRB,Imai_superconductor_1995,JAFL_PRL_BaSi2,Evers_Ortho74,Demchyna_Kristall,Yamanaka_sp3_chem}. 
Hence GICs and doped graphene are not unique systems, having a Si and Ge counterpart can be seen as members of a generalized family of group-IV intercalated honeycomb lattices (gIV-ICs). 

So far the highest \tc\ reported on gIV-ICs is 11.5\,K in \cac~\cite{Weller_CaC6_NAT2005,Nemery_PRL_CaC6}. 
This system is also the most studied among the family and its superconducting properties are rather well 
understood~\cite{MCalandra_CaC6_PRL,Boeri_PRB_CaC6,Boeri_CaC6-pressure,Sanna_CaC6_RapCom2007,Yang_Nat2014}. 
It is particularly clear that an important role is played by the existence at the Fermi level of 2D electron like bands 
as well as anti-bonding C-$\pi$ states. It is also know that a sufficiently large intercalation is therefore a necessary 
condition to obtain high critical temperatures. 
But what is the highest conceivable \tc\ in an intercalated graphite-like system? Could Si and Ge iso-morphs be better candidates than GICs? We will address these questions by focusing our investigation on the high doping limit, 
with one intercalating atom per two honeycomb atoms. 
We will indicate this family of compounds as \mx\ where M stands for a metal of the I and II column of the periodic table 
and X is carbon, silicon or germanium. This composition is known to occur~\cite{Demchyna_Kristall} in several silicides~\cite{Evers_Ortho74,Bordet_PRB_CaSi2,JAFL_PRL_BaSi2} and germanides~\cite{Tobash_CaGe2_JSSChem2007,Zintl_EuGe2_JSSChem2004,Yamanaka_sp3_chem}. 

We will show by means of theoretical \textit{ab-initio} methods, that finding high temperature superconductivity 
in these families is a false hope. On the other hand breaking the record critical temperature of \cac\ is likely to be possible. 

\begin{figure}[t]
\includegraphics[width=1.0\columnwidth,angle=0]{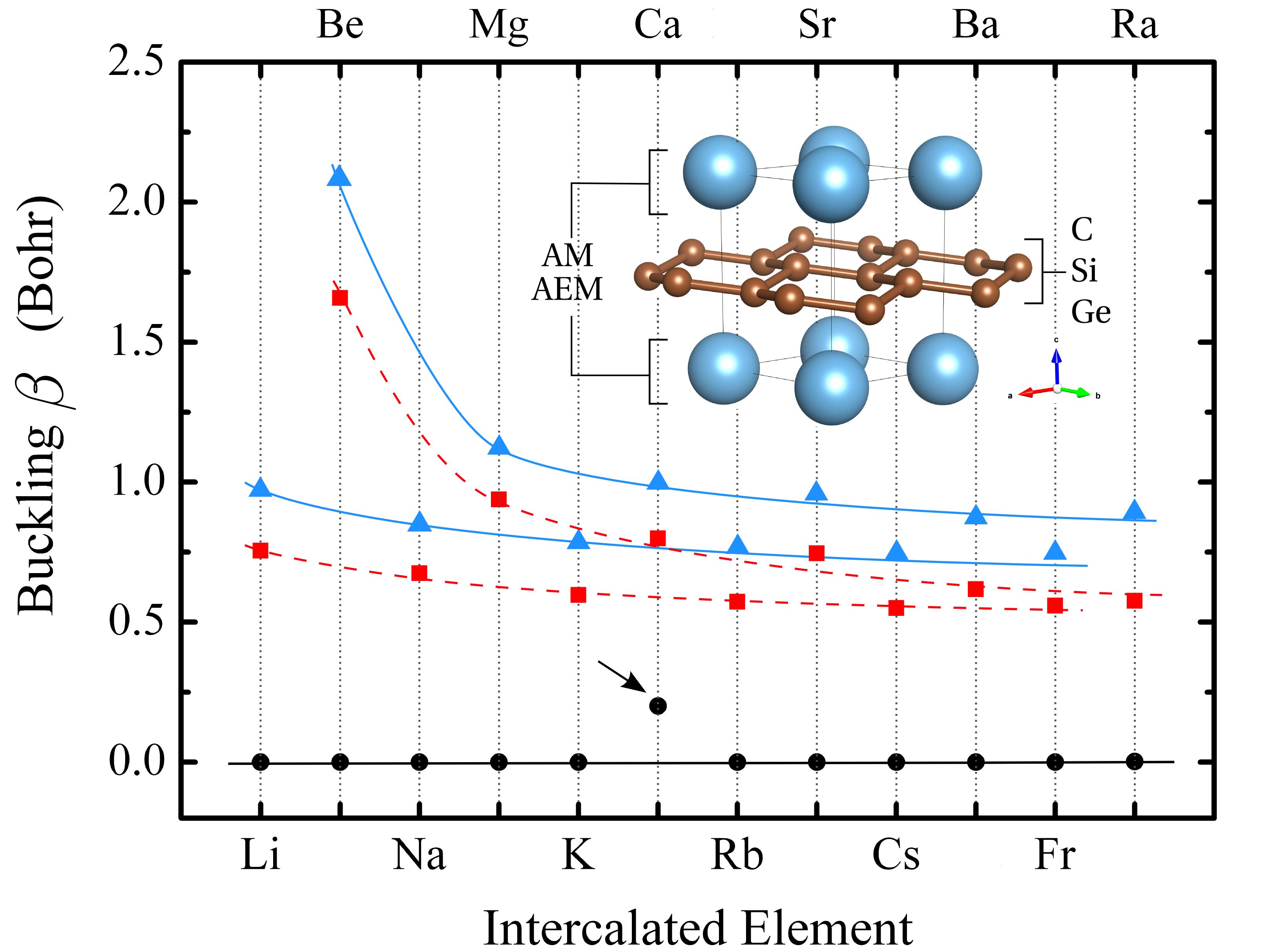}
  \caption{(Color online) Buckling ($\beta$) of the honeycomb layer as a function of 
  the chemical composition from theoretical structural relaxation. 
  Intercalated graphites are shown as black dots, silicides as red squares and 
  germanides as blue triangles. Lines are guide to the eye to stress 
  the different behavior of alkali and alkaline earth intercalation. 
  The inset shows a prototype crystal structure in a buckled configuration ($\beta\ne0$, EuGe$_2$ crystal type).}
 \label{fig:zplot}
\end{figure}

All systems are structurally relaxed within Kohn-Sham density-functional theory. 
\footnote{We used the two plane-wave based code {\sc abinit}~\cite{gonze_abinit_2009}, and {\sc espresso}~\cite{QE-2009} 
 within the Perdew-Burke-Ernzerhof (PBE)~\cite{GGA-PBE} exchange correlation functional and the core 
states were accounted for by norm-conserving Troullier-Martins pseudopotentials~\cite{FHI_Fuchs}.
The pseudopotential accuracy has been checked against all-electron (LAPW+lo) method as implemented in the {\sc elk} code (http://elk.sourceforge.net/).}
Upon relaxation~\cite{sup_mat} all carbon compounds, apart from CaC$_2$, converged to the AlB$_2$ crystal structure (space group $P6/mmm$, number 191), while all silicides and germanides as well as CaC$_2$ converged to the EuGe$_2$ crystal structure (space group $P\bar{3}m1$, number 164). 
In both, M occupies the $1a$ Wyckoff position (0,0,0) and X the the $2d$ positions (1/3, 2/3, $z$) and (2/3, 1/3, $-z$). 
In the  AlB$_2$ structural prototype the parameter $z$ is fix to $1/2$, while in the EuGe$_2$ structure it is related to a buckling ($\beta$) of the honeycomb lattice: $\beta = (|z-1/2| \cdot c)$. 
The EuGe$_2$ structural prototype and the values of $\beta$ are shown in Fig.~\ref{fig:zplot}. 
This figure shows clearly that intercalating lighter ions (Li, Be, Na) induce high buckled honeycomb plans, 
while heavier ions (Rb,Cs,Ba) tend to induce low-buckled plans. 
CaC$_2$ deviates from the general trend, this structure has a mixture of $sp^2-sp^3$ (75$\%-25\%$ respectively) bonding  
and therefore at ambient pressure it present a finite buckling (energetically more favorable than in a flat AlB$_2$ structure).  
In this respect, it has been recently predicted by Li and coworkers~\cite{LiYangLing_PressCaC2_PNAS2013} that the 
flat-layered phase could be stabilized at high-pressures. 

As many of the compounds discussed in this work are not experimentally known,  in order to assert on 
their potential synthesis we calculated their thermodynamic stability, 
this is derived from the total DFT energy of the system (\mx\ )  and  of its elemental ground state solid 
(see supplemental  material \cite{sup_mat} for details). This analysis leads to the conclusion that all 
graphite compounds in the \mx\ layered phase are unstable towards this elemental decomposition. 
While most of silicides and germanides are stable towards decomposition. Nevertheless, since a positive formation energy 
does not completely exclude these materials from their possible synthesis, we will also investigate their 
dynamical stability (phonons).

For all systems under investigation we computed phonons and only for those systems dynamically stable, the 
electron-phonon coupling was calculated by means of density-functional perturbation theory.~\footnote{The 
phonon spectrum and the electron-phonon matrix elements were obtained employing density-functional 
perturbation theory.~\cite{dfpt_s.baroni,2n+1_gonzepaper,savrasov2_prb}, within the pseudopotential approximation. 
A cutoff energy of 60\,Ry was used in the plane-wave expansion. A $16\times16\times12$ Monkhorst-Pack $k$-grid and a $4\times4\times2$ $q$-grid was used for all the materials under consideration. 
With the only exception of \src, \rs\ and \rg, where we have increased the $q$ sampling grid to 
$8\times8\times6$ in order to achieve an accurate description of anisotropic properties.} 
We found most of the intercalated carbon compounds to be dynamically unstable, with the only exception of Sr and Ca intercalation. 
This suggests that the 1 to 2 intercalation is too large for this family and is evidenced experimentally by 
the reported challenging synthesis of LiC$_2$~\cite{Belash_LiC2_SSC1989}, that turns out to be metastable, 
partially loosing its Li content and converting in LiC$_6$~\cite{Nalimova_LiGICs_Carbon1995,Avdeev_AlkaliGICs_HiPRes1990}.
On the other hand, with the exception of light-ion intercalants, most of the disilicides and digermanides are dynamically stable.  

\begin{figure}[t]
\includegraphics[width=0.96\columnwidth,angle=0]{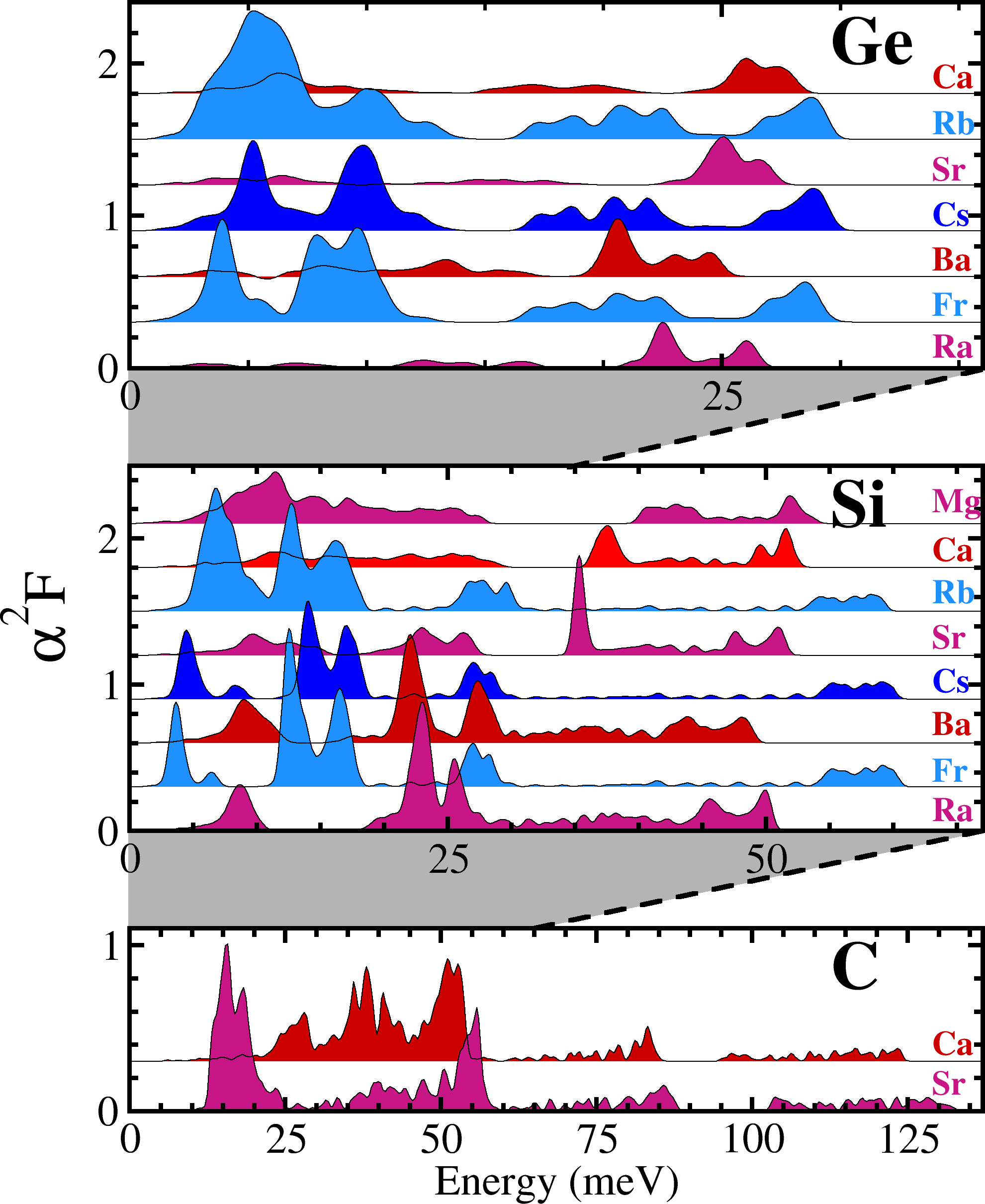}
  \caption{(Color online) 
 Eliashberg spectral function $\alpha^2F(\omega)$ calculated for dynamically stable materials on the \mx\ set under investigation. 
 These spectral functions show an overall similar behavior, having:  
 a low energy region dominated by intercalant phonon modes, 
 a middle energy range with out-of-plane $X$ phonon modes and 
 an high energy spectra of $X$-bonds stretching modes. 
 These features are also seen in \cac\ and doped graphene~\cite{MCalandra_CaC6_PRL,Profeta_Graphene_Nat2012}. 
 Note that the plots have different frequency scale, as the mass of the honeycomb atoms 
 scales the entire spectral function and for both: low and high energy regions.}
 \label{fig:a2f}
\end{figure}

Eliashberg spectral functions~\cite{Carbotte_RMP1990,AllenMitrovic1983} $\alpha^2F(\omega)$ for all the dynamically stable systems are reported in Fig.~\ref{fig:a2f}. From now on we will only consider this subset of materials. 
In this figure we can clearly observe that the spectral functions are scaled in their frequency 
by the mass of the atom forming the honeycomb layer. And this extends not only, obviously, to the high 
energy modes that originate from strong in-layer bonds, but also to the low frequency 
modes that are dominated by the intercalant motion in the weak interlayer potential. Thus, indicating a chemical effect.  
We also observe that alkali metals (as compared with alkaline earths) lead to systematically lower phonon branches, 
therefore to an enhanced coupling strengths~\cite{Carbotte_RMP1990}

\begin{equation}\label{eq:lambda}
\lambda=2\int\frac{\alpha^2F(\omega)}{\omega}d\omega,
\end{equation}

at the same time this lowers the average frequency, that we conventionally express as 

\begin{equation}\label{eq:wlog}
\omega_{\rm log}=\exp\left[\frac{2}{\lambda}\int\alpha^2F(\omega)\frac{\ln(\omega)}{\omega}d\omega \right].
\end{equation}

From an electronic point of view, all the materials share a qualitative similar structure.
As in the case of CaC$_6$ or doped graphene, there are two type of electronic states located at the Fermi energy: 
anti-bonding $\pi$ states provided by the honeycomb layer (C, Si, Ge) and 2D interlayer states with contributions 
from the M $d$-orbitals. These electronic states hybridize differently along the alkali or the alkaline-earth column 
and lead to different effective doping and band alignment. This affects the density of states at the Fermi energy 
(\dos) and whit it the occurrence of superconductivity, as we will show below. 

\begin{figure}[t]
\includegraphics[width=0.95\columnwidth,angle=0]{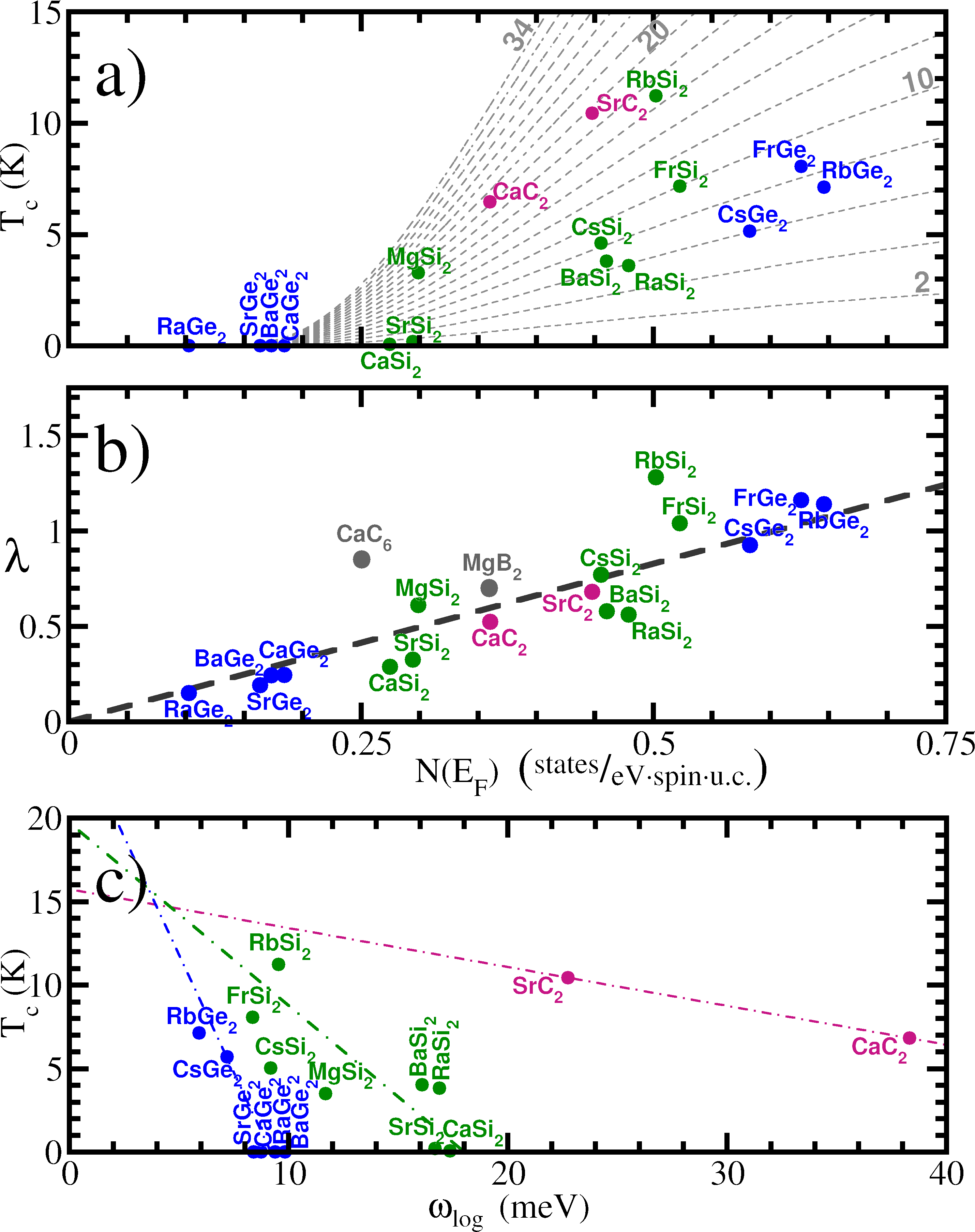}
\caption{(Color online) 
a) Critical temperatures calculated with within the McMillan-Allen Dynes 
formula~\cite{McMillanTC,AllenDynes_PRB1975} using $\mu^*=0.1$. 
Grey dashed lines are isomass lines representing  the phonon energy average \omlog\ (Eq.~\ref{eq:wlog}) 
ranging from 2 to 34~\,meV; b) Electron-phonon coupling parameter $\lambda$ (Eq.~\ref{eq:lambda}) as a function of 
the density of state at the Fermi level ($N(E_F)$); c) Critical temperature versus \omlog (lines serves only as guide).} 
 \label{fig:lambda}
\end{figure}

In order to perform a fast screen  of our \mx\ set, the superconducting critical temperatures were estimated within 
McMillan-Allen-Dynes parametrization of the Eliashberg equations~\cite{McMillanTC,AllenDynes_PRB1975,Carbotte_RMP1990,Eliashberg} 

\begin{equation}\label{eq:mcmillan}
  T_\text{c}=\frac{\omega_\text{log}}{1.2 k_B}\exp\left[-\frac{1.04(1+\lambda)}{\lambda-\mu^{*}(1+0.62\lambda)}\right], \\
  \end{equation}
  
where $k_B$ is the Boltzmann constant. This formula depends on three parameters:
the Coulomb pseudopotential \mus\ (here fixed to $0.1$ by comparison with SCDFT results, see below); the logarithmic 
average of the phonon frequency \omlog; and the coupling constant $\lambda$. 
The computed \tc\, couplings $\lambda$ and \omlog\ are shown in Fig.~\ref{fig:lambda}.  

In the limit of an homogeneous coupling in {\bf k}-space, $\lambda$ is proportional to \dos.  
Within BCS theory, this parameter splits as $\lambda = V$\dos, where $V$ is the BCS coupling strength. 
In Fig.~\ref{fig:lambda}~b) we observe a remarkable proportionality between $\lambda$ and \dos. 
Leading to the conclusion that $V$ is approximately the same on this \mx\ class of systems, with the sole exception 
of few systems characterized by strong softening. Eventually this softening will leads to a phononic instability 
and to a structural phase transition. Perhaps under different thermodynamic conditions of pressure and temperature. 
Although not belonging to this \mx\ family we observe that \cac\ lies perfectly in 
this regime~\cite{Sanna_CaC6_RapCom2007,Sanna_CaC6ELI_PRB2012}.  And similarly does MgB$_2$, however, 
this is accidental as we have ignored its multi-band nature~\cite{Mazin_MgB2_PRL2001,Floris_MgB2_PRL2005,Floris_MgB2_PhysicaC2007}.

These calculations predict several interesting superconductors and in particular \rs,  \rg\ and \src. 
\rg\ has the highest density of states and, as discussed above, also presents the highest $\lambda$, even though it 
shows a modest \tc\ of 7\,K. In fact \tc\ (see Eq.~\ref{eq:mcmillan}) depends also on the phonon energy, 
which is larger for systems having lower mass, for instance \src. Also in this Fig.~\ref{fig:lambda} (in panel a), 
we included the iso-mass lines as a reference to indicate how the \tc\ in a material would be affected 
by $\lambda$ (on $X_2$) or by \dos. 
The outcome of this analysis suggests the existence of an upper critical temperature for each family. 
And this is imposed by the electronic structure, as \dos\ hardly would exceed the value of  0.7\,states/eV/spin (of \rg ).  
Following the iso-mass lines in this figure for each subfamily, leads to the conclusion that an upper critical 
temperature of about 10\,K, 15\,K and 20\,K exists respectively for intercalations 
in Ge, Si and carbon honeycombs. We firmly believe that this conclusion can be extended beyond the \mx\ class, 
since different intercalation density will not plausibly affect the coupling strength. 
However, the coupling strength could be significantly affected if $\sigma$ states were involved (as in MgB$_2$), 
but this would require an unphysical doping level.

We will now focus our investigation on three selected systems \src, \rs\ and \rg\ as the most interesting 
representative of each sub-family. 
As discussed in the introduction both \rs\ and \rg\ are stable towards elemental decomposition. 
In addition they are also more stable than their RbSi$_6$ and RbGe$_6$ rhombohedral counterparts~\cite{sup_mat}. 
Therefore, we believe, these two systems are likely to be accessible to the experimental synthesis. 
On the other hand, \src\ is not stable with respect to elemental decomposition and turn to be less energetically 
competitive than its rhombohedral SrC$_6$ configuration that, in fact, has been synthesized~\cite{Kim_Boeri_SrC2}. 
Nevertheless, since the system is dynamically stable, it may still be possible to find a way to its synthesis, 
perhaps by means of a non-equilibrium process  or by high temperature and high pressure, as often used to synthesize  clathrates,~\cite{Toulemonde_PRB} carbon borides,~\cite{CaB_phases_HP} and layered disilicides~\cite{Evers_Ortho74,MImai_ChemMat,Bordet_PRB_CaSi2,JAFL_PRL_BaSi2} and germanides~\cite{Tobash_CaGe2_JSSChem2007}. 

\begin{figure}[t]
\begin{center}
\includegraphics[width=0.79\columnwidth,angle=0]{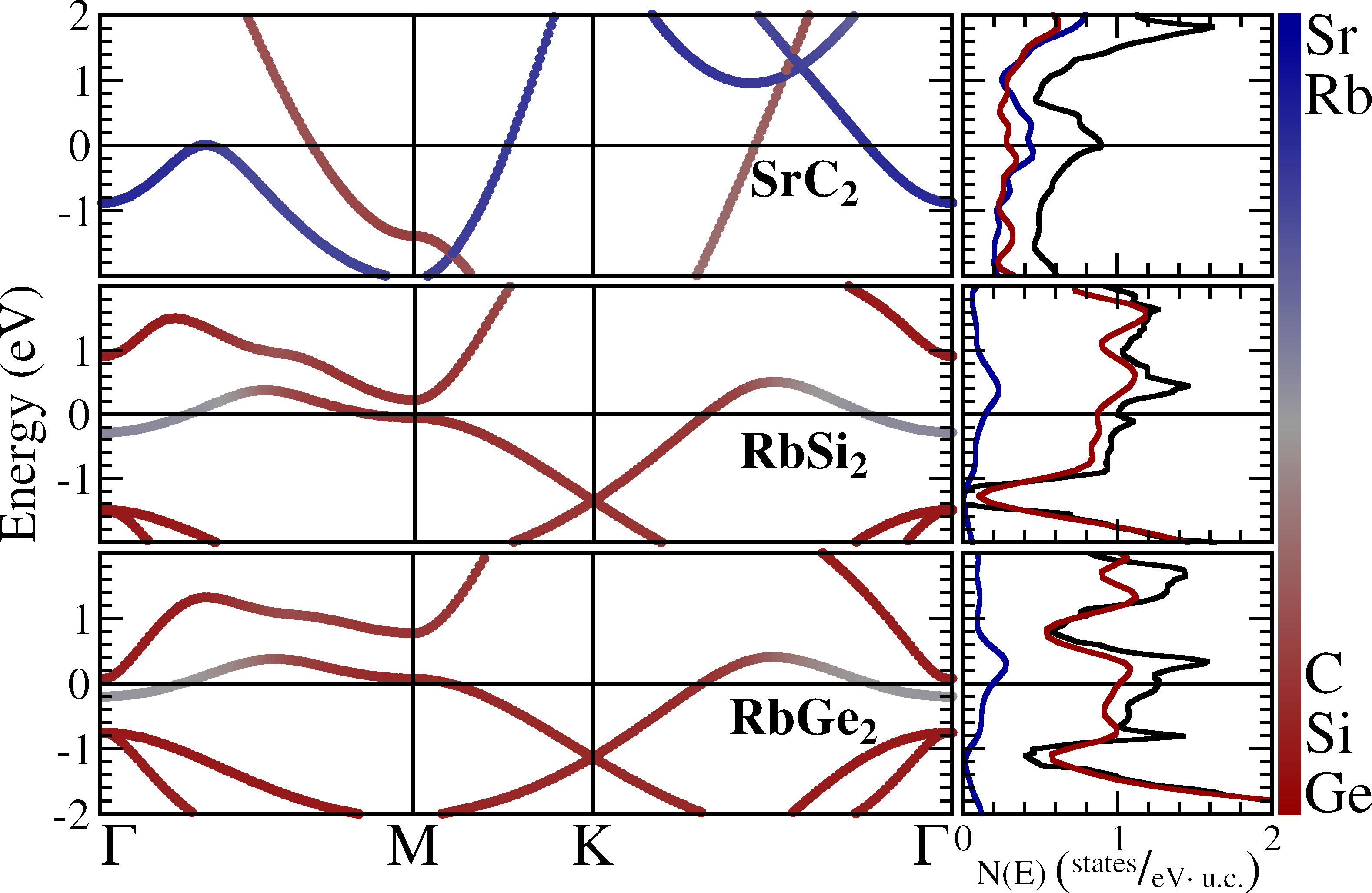}
\vspace{0.4cm}
\includegraphics[width=0.80\columnwidth,angle=0]{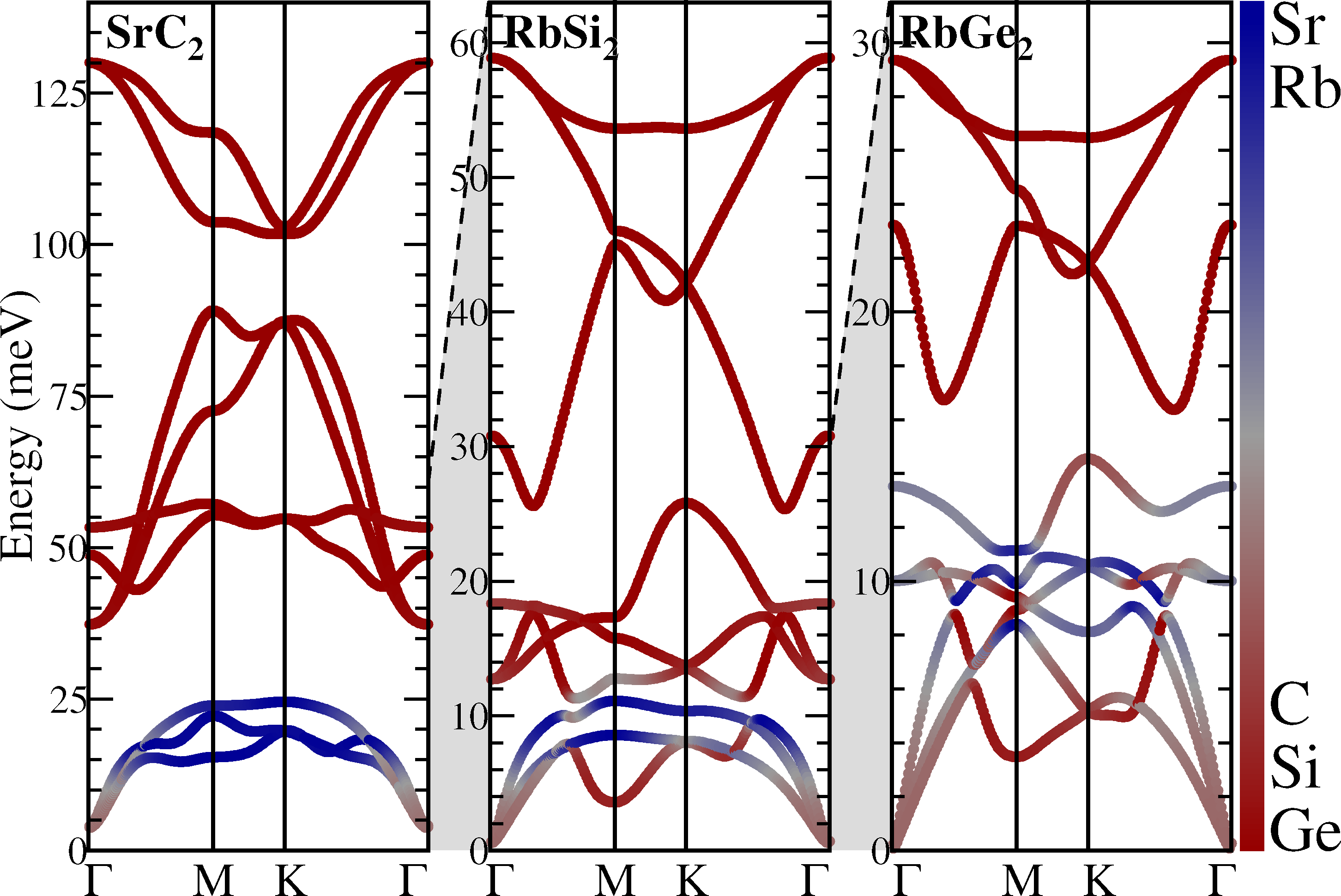}
\caption{(Color online) Top: Electronic bands in \src, \rs\ and \rg\ around the Fermi level (at zero eV). 
 The color scale indicates the projection of the Kohn-Sham states on the atomic orbitals of the intercalating atom. 
 Bottom. Phonon dispersion relation. The color scale indicates the component of the phonon mode on the intercalant atom.} 
 \label{fig:bands_and_phonons}
\end{center}
\end{figure}

\begin{figure}[t]
\includegraphics[width=0.96\columnwidth,angle=0]{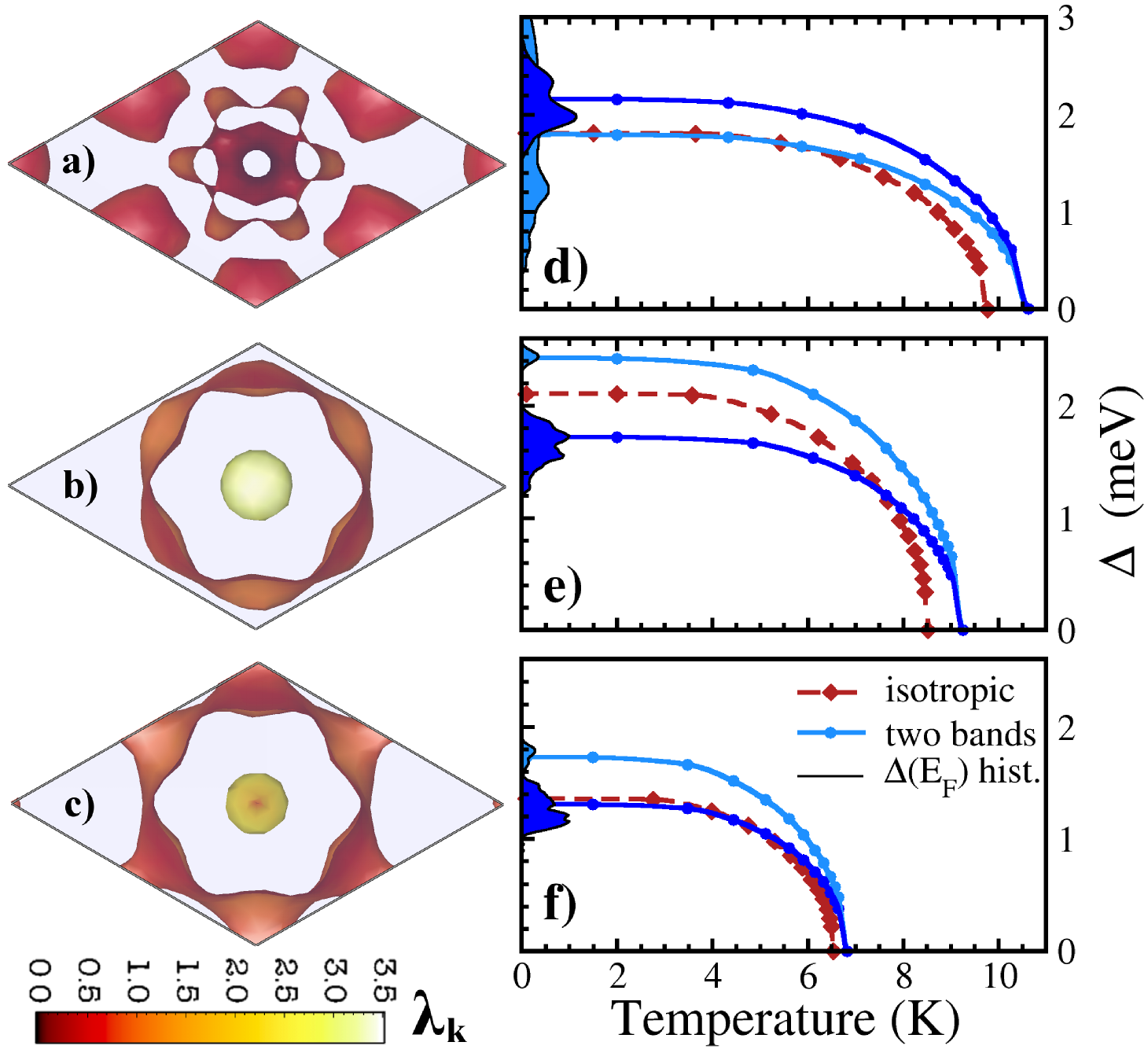}
\caption{(Color online) Fermi surface of \src\ (a) \rs\ (b) and \rg\ (c), 
shown in the $\Gamma$ centered reciprocal unit cell (top view). 
The color scale (bottom left corner) gives the ${\bf k}$-resolved electron phonon
coupling~\cite{Sanna_CaC6_RapCom2007} $\lambda_k$. 
Superconducting gap as a function of temperature for \src\ (d) \rs\ (e) and \rg\ (f), 
computed within SCDFT~\cite{Sanna_Migdal}. 
The red-dashed line is the isotropic behavior, blue-continuous lines are a minimal two-band approximation. 
The full gap distribution function~\cite{Sanna_CaC6_RapCom2007,Sanna_CaC6ELI_PRB2012} is given at $T=0$ in as a filled area.}
 \label{fig:scdft}
\end{figure}

The electronic band structures of these three selected materials are shown in Fig.~\ref{fig:bands_and_phonons}. 
The bands of \src\ essentially differ from those of \rs\ and \rg, due to the effect of symmetry breaking (both have buckling) 
and as well the doping level of the honeycomb lattice (charge projection shows that divalent strontium donates $1.2$ 
electrons - while monovalent Rb donates $0.5$ electrons for both \rs\ and \rg). 
The Fermi surfaces (FS) shown in Fig.~\ref{fig:scdft} a, b and c, present multiple Fermi sheets with different orbital character. 
In \src\ the inner FS comes from interlayer states, while the outer surface is formed by carbon $\pi$ states. 
In \rs\ and \rg\ the hybridization between interlayer and honeycomb $\pi$ states is much stronger. The outer FS is mostly due to Si/Ge $\pi$ states, while the inner FS has an interlayer character, however with a relatively large overlap (25\%) to Si/Ge $\pi$  states. 

The phonon dispersion for the three systems is shown in Fig.~\ref{fig:bands_and_phonons}. 
The overall structure of the phonon modes is the same for the three systems. 
Low frequency modes present a strong intercalant component, fundamentally due to the weak force constants 
that binds the $M$ atoms to their position in the lattice, but also because of their relatively large mass. 
To the scope of this work, the most interesting feature of the phononic dispersion is the behavior of the buckling modes.  
In the unbuckled (flat) \src\ compound this mode has 50\,meV in the zone center and and cannot falls below 40\,meV. 
While in the buckled \rs\ and  \rg\ compounds it becomes ``soft'' moving from $\Gamma$ ( at 30\,meV in \rs\ and 23\,meV in \rg) to $M$ (3.5\,meV). This mode is strongly coupled in both \rs\ and \rg, and anharmonic effects (not considered in the present work) 
may also affect the strength of its coupling.

We will now reconsider the superconducting properties of these selected systems 
by means of a more accurate superconductivity theory than the McMillan formula used so far. 
We will adopt density-functional theory for superconductors (SCDFT), as it is completely parameter free~\cite{OGK_SCDFT_PRL1988,Lueders_SCDFT_PRB2005,Marques_SCDFT_PRB2005} and allows for a full 
{\bf k}-resolved description~\footnote{The phononic functional we use is an improved version with respect to Ref.~\onlinecite{Lueders_SCDFT_PRB2005,Marques_SCDFT_PRB2005} 
and is discussed in Ref.~\onlinecite{Sanna_Migdal}. 
Coulomb interactions are included within static RPA~\cite{Sanna_CaC6_RapCom2007}.}

It should be observed (see Fig.~\ref{fig:scdft} on panels a, b and c) that the electron phonon coupling in all these systems is rather anisotropic, meaning strongly {\bf k}-dependent on the FS. \src\ 
has a continuous distribution, while the two FS of \rs\ and \rg, have remarkably 
different coupling strength:  stronger on the small FS around the $\Gamma$ point and weaker 
in the outer FS (at large |{\bf k}|). The distribution of superconducting gaps on the Fermi 
energy (not shown) follows the anisotropy 
in $\lambda_{\bf k}$, similarly to the behavior observed in bulk lead~\cite{Floris_Pb_PRB2007}.
The gap distribution function at $T=0$ (i.e. the  energy distribution of the SC gaps: $\Delta_{k_F}$), as well as the temperature dependence 
(in a two-band and single band model) are plotted in Fig.~\ref{fig:scdft} d,e,f. Both \rs\ and \rg\ show two distinct gaps 
(like in MgB$_2$ or bulk lead~\cite{Floris_Pb_PRB2007,Floris_MgB2_PRL2005,Floris_MgB2_PhysicaC2007}), while \src\ has an anisotropic gap continuously distributed. 
This gap distribution reminds that of \cac~\cite{Sanna_CaC6_RapCom2007,Gonnelli_CaC6_PRL2008,Nagel_CaC6_PRB2008}. 
This anisotropy will affect the specific heat and the thermodynamical properties. However, unlike in MgB$_2$, the critical temperature 
is not much affected by it (less than 1\,K). The role of coupling anisotropy on the superconducting behavior can be clearly 
understood within the qualitative model of Suhl, Mattias and Walker~\cite{SMW_multibandBCS_PRL1959}. 
The observed combination of a large anisotropy in the gap with a small enhancement in \tc\ is a consequence of the 
strong inter-band coupling between $\pi$-states (having a smaller gap) and the interlayer states (that dominate on the larger gap). 
The gap distribution of \src\ is even broader and clearly cannot be completely captured within a two-band model. 
The system is in fact almost gapless, since the small $|{\bf k}|$ part of the FS shows a negligible superconducting pairing as a consequence of the weak phononic coupling.

In summary, we presented a theoretical study on honeycomb layered binary carbides, silicides and germanides 
intercalated by alkali or alkaline-earth metals. Our superconductivity analysis has shown that in this class of 
materials are many compounds with a relatively high critical temperature ($\sim$ 10\,K) as well as a quite complex 
superconducting state. In addition, the stability investigation has shown that several compounds should be 
accessible to their experimental synthesis. Finally, we demonstrate an intrinsic physical similarity among the group, 
which can be traced back to their characteristic $\pi$+interlayer character of states at the 
Fermi surface. From this feature we estimate an upper limit for the transition critical temperatures: 
$\sim20$\,K, $\sim15$\,K and $\sim10$\,K respectively for carbon, silicon and germanium intercalated honeycombs. 
This limit could be broken only in the unlikely case in which the doping level would be able to drive $\sigma$-states 
at the Fermi level. Nevertheless this study indicates that superconductivity in doped graphite and similar systems 
is a rather general behavior and many more superconductors may still be discovered. 

J.A.F.L. acknowledge financial support from EU's 7th Framework Marie-Curie scholarship Program within the ``ExMaMa'' Project (329386).  

\bibliographystyle{apsrev4-1}
\bibliography{paper}
\end{document}